\documentclass{article}
\usepackage[utf8]{inputenc}
\usepackage[english]{babel}
\usepackage{amsmath}
\usepackage{comment}
\usepackage{graphicx}
\usepackage[colorlinks,hypertexnames=false]{hyperref}
\usepackage{cite}
\graphicspath{ {./images/} }
\usepackage{indentfirst}
\usepackage{float}
\usepackage{tikz}
\usepackage{gensymb}

\title{Internal kinematics of dwarf galaxies orbitally moving in ultralight dark matter}
\author{E.V.~Gorbar${}^{1,2}$ and A.I. Momot${}^{1}$\\
${}^1$ \it \small Faculty of Physics, Taras Shevchenko National University of Kyiv,\\
\it \small 64, Volodymyrs'ka str., Kyiv 01601, Ukraine\\
${}^2$ \it \small Bogolyubov Institute for Theoretical Physics, National Academy of Sciences of Ukraine,\\
\it \small 14-b Metrolohichna str., Kyiv 03143, Ukraine
}
\date{}

\begin{document}

\maketitle

\section*{Abstract}

For dwarf galaxies modeled as deformed Plummer spheres and orbitally moving in ultralight dark matter halo of the Milky Way, the torque induced by the dynamical friction force is determined. The impact of this torque, as well as the torque produced by the gravitational force of the Milky Way, on the internal kinematics of dwarf galaxies is studied. Possible oscillations of dwarf galaxies caused by the tidal torque and misalignment of dwarf galaxies with respect to their equilibrium position are investigated and the corresponding frequencies and periods of oscillations are found.

\section{Introduction}
\label{sec:introduction}

Dwarf galaxies are small galaxies, typically containing a few million to a few billion stars, compared to hundreds of billions in large galaxies like the Milky Way. The mass-to-light ratio in dwarf galaxies can exceed 100 (solar units) compared to 10-20 in normal galaxies. In addition, the velocity dispersion of stars in dwarf galaxies is larger than expected based on the visible matter alone. This suggests that dwarfs are strongly dark matter dominated structures which makes them excellent laboratories for the study of dark matter and its properties \cite{Lelli,Sales,Boldrini}.

Dwarf galaxies are mainly observed in the Local Group orbiting the Milky Way and Andromeda \cite{McConnachie}. There are also 61 dwarf galaxies confirmed to be within 420 kiloparsecs of the Milky Way \cite{Simon}. 
The orbits of these dwarf galaxies lie far from the Milky Way center, where dark matter begins to dominate over baryonic matter as the Milky Way rotation curve implies (see, e.g., \cite{Schneider}). Indeed, while the Milky Way diameter is around 27 kpc, cosmological simulations suggest that the edge of the Milky Way halo is situated at 292 kpc \cite{Navarro}. The visible disk of the Milky Way Galaxy is thought to be embedded in a much larger, roughly spherical halo of dark matter whose density decreases with distance from the galactic center. This means that for dwarf galaxies orbiting the Milky Way, the dynamical friction force due to dark matter should dominate compared to the contribution due of baryonic matter.

Models of ultralight dark matter (ULDM) with boson particle masses in the range of $10^{-23}$ to $10^{-21}$ eV attract considerable interest due to their unique phenomenological features and have been the subject of extensive research (for a review, see Refs.\cite{Ferreira,Calmet,Matos}). These models reproduce the large-scale structure of the Universe similarly to the cold dark matter (CDM) model and, in view of the ULDM kpc scale of the de Broglie wavelength, could address several issues that CDM faces on the galactic scale.

Dynamical friction experienced by moving stars, globular clusters, and dwarf galaxies is a classical subject \cite{Chandrasekhar} which is very important in the study of the kinetics of astronomical bodies. In ULDM models, the dynamical friction was studied for point-like probes in the fuzzy ultralight dark matter (no self-interactions among dark matter particles are absent) in \cite{Ostriker,Lancaster,Wang,Boey}. The role of ULDM self-interaction for the frictional force acting on stellar bodies was investigated in \cite{Desjacques,Boudon,Hartman,Buehler,Berezhiani}. Note that, using the radial distribution of stars in six UFDs measured by the Hubble Space Telescope, it was recently argued \cite{Almeida} that collisions among dark matter particles, i.e., self-interaction is necessary to explain the observed stellar distribution.

Globular clusters and dwarf galaxies are spatially extended objects which are often modeled as Plummer spheres. This motivated the study of the dynamical friction force acting on circularly moving Plummer spheres performed in \cite{Gorkavenko,Barabash}, where analytic expressions for both radial and tangential components of the dynamical friction force were found.

In ULDM models, the dark matter profile in galaxies is characterized by the presence of a dense solitonic core in the state of a Bose-Einstein condensate, surrounded by a diffuse halo. Since dwarf galaxies are dark matter dominated, the baryonic matter (stars and gas) distribution follows that of dark matter. It was shown in \cite{Pozo} that the stellar density distribution (and the dark matter distribution as well) can be reasonably well approximated within the core by a Plummer profile. We use this profile in our study here to model the matter distribution in dwarf galaxies aiming to address their internal kinematics caused by the dynamical friction and tidal forces.

An important parameter in the description of ultralight dark matter far from the solitonic core is an effective temperature produced by gravitational cooling and violent relaxation, which explains how effectively collionless self-gravitating systems acquire an effective temperature \cite{chavanis2019predictive,Launhardt,Schonrich,Portail,Schive,Chavanis:2022}. Adding phenomenologically temperature to the Gross-Pitaevskii equation as a parameter in a logarithmic nonlinearity (accounting for an “effective thermal pressure”) produces ULDM halos with central solitonic cores and outer isothermal atmospheres \cite{chavanis2019predictive,Chavanis:2022}. The latter provides the corresponding environment where dwarf galaxies orbit the Milky Way in ULDM models.

The paper is organized as follows. Our model is described and the dynamical friction force for an orbiting deformed Plummer sphere is derived in Sec.~\ref{sec:model}. The torque induced by the dynamical friction force is calculated in Sec.~\ref{sec:torque}. The torque due to tidal forces acting on orbiting dwarf galaxies is determined in Sec.~\ref{sec:torque-tidal}. Numerical results for torques are presented in Sec.~\ref{sec:numerical}. Oscillations of dwarf galaxies with respect to their equilibrium position are studied in Sec.~\ref{sec:period}. The results are discussed and summarized in Sec.~\ref{sec:Conclusion}.

\section{Model and dynamical friction force}
\label{sec:model}
\vspace{8mm}

Let us consider a dwarf galaxy of mass $M$ which moves on a circular orbit of radius $r_0$ with constant angular velocity $\Omega$ in the steady-state regime. Although dwarf galaxies are often modeled \cite{Irwin,McConnachie,Lokas,Walker} as spherically symmetric Plummer spheres, the tidal and dynamical friction forces may deform dwarf galaxies during their orbital motion. In fact, ultra-faint dwarf galaxies are rather elongated systems \cite{Martin,Revaz} with a mean ellipticity of about $\varepsilon=0.4$.

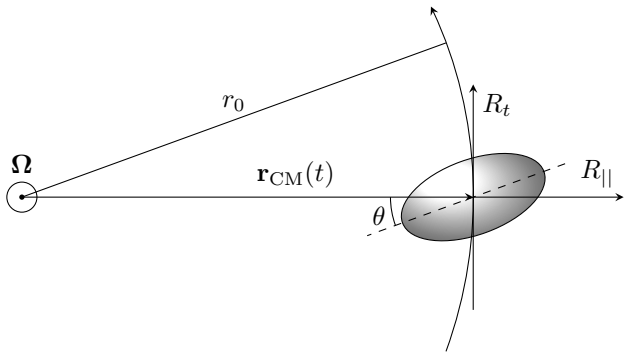
\begin{figure}[htb]
	\centering
\begin{tikzpicture}[>=stealth,scale=1]
	\shade[ball color=white,rotate around={20:(0,0)}] (0,0) ellipse [x radius=1cm, y radius=0.5cm];
	\draw[rotate around={20:(0,0)}] (0,0) ellipse [x radius=1cm, y radius=0.5cm];
	\draw[->] (-6, 0) -- node[anchor=south west] {$\mathbf{r}_\text{CM}(t)$} (0, 0) ;
	\draw[->] (0, 0) -- (2, 0) node[anchor=south east] {$R_{||}$};
	\draw[->] (0, -1.5)  -- (0, 1.5) node[anchor=north west] {$R_t$};
	\draw[dashed, rotate around={20:(0,0)}] (1.3,0) -- (-1.5,0);
	\draw (-1.1,0) arc[start angle=-180, end angle=-160, radius=1.1cm];
	\draw (-1.25,0) node[anchor=north] {$\theta$};
    \draw[->,rotate around={-20:(-6,0)}] (0,0) arc (0:45:6cm);
    \draw[rotate around={20:(-6,0)}] (-6,0)  -- node[anchor=south] {$r_0$} (0,0);
    \draw (-6,0) circle[radius=0.2cm] ;
    \fill (-6,0) circle[radius=1pt] node[above=2mm]{$\mathbf{\Omega}$};
\end{tikzpicture}
\caption{Schematics of the plane of an orbiting dwarf galaxy modeled as a deformed Plummer sphere.
\label{scheme}}
\end{figure}

Without loss of generality, we could assume that the circular orbital motion of a dwarf galaxy takes place in a plane with radius $r_0$ so that the radius vector of its center of mass is given by
$$
\mathbf{r}_\text{CM}(t)=(r_0\cos(\Omega t),r_0\sin(\Omega t),0).
$$
As to the deformed Plummer sphere that we use to model the dwarf galaxy, a natural assumption is to consider it as an ellipsoid elongated along $\mathbf{r}_\text{CM}(t)$ due to tidal forces with the largest axis of the ellipsoid rotated by angle $\theta$ with respect to $\mathbf{r}_\text{CM}(t)$. Then, the deformed Plummer sphere density profile is defined by
\begin{equation}
\rho(\mathbf{r}-\mathbf{r}_\text{CM}(t))=\frac{3M}{4\pi l^3_Pab}\frac{1}{\left(1+\frac{R^2_z}{l^2_p}+\frac{(R_{||}\cos\theta+R_t\sin\theta)^2}{a^2l^2_P}+\frac{(-R_{||}\sin\theta+R_t\cos\theta)^2}{b^2l^2_P}\right)^{5/2}},
\label{Plummer-sphere-deformed}
\end{equation}
where $a$ and $b$ are the deformation parameters. Notice that, for $\theta=0$ and $a=b=1$, the density profile (\ref{Plummer-sphere-deformed}) reduces to the standard Plummer sphere profile of radius $l_p$. In addition, we assume that $a>1$, $b<1$, and $\theta \ll 1$ so that dwarfs are elongated mostly along $\mathbf{r}_\text{CM}(t)$ due to tidal forces with small inclination in the tangential direction and are flattened in the tangential direction due to the dynamical friction force. Further, $R_z\equiv z$, $R_{||}(t)=(\mathbf{r}-\mathbf{r}_\text{CM}(t))_{||}$ and  $R_t(t)=(\mathbf{r}-\mathbf{r}_\text{CM}(t))_t$ are the projections of $\mathbf{r}-\mathbf{r}_\text{CM}(t)$ onto $\mathbf{r}_\text{CM}(t)$ and the tangential direction $\dot{\mathbf{r}}_\text{CM}(t)/|\dot{\mathbf{r}}_\text{CM}(t)|$, respectively. Finally, we introduced the factor $1/ab$ in Eq.(\ref{Plummer-sphere-deformed}) so that the deformed Plummer sphere has the same mass $M$ as nondeformed Plummer sphere. Geometrically, the considered deformed Plummer sphere is a prolate spheroid rotated by angle $\theta$ with respect to $\mathbf{r}_\text{CM}(t)$ (see Fig.~\ref{scheme}).

In our derivation of the dynamical friction force, we follow the setup developed in \cite{Gorkavenko} based on the approach considered in \cite{Desjacques,Buehler}.
A deformed Plummer sphere moving in homogeneous ultralight dark matter with density profile $n_0(r)$ composed of ultralight bosonic particles of mass $m$ perturbs due to gravitational interaction the ULDM density $n_\text{DM}(t,\mathbf{r})=n_0(r)(1+\alpha(t,\mathbf{r}))$. In the linear response approach, the corresponding density inhomogeneity $\alpha(t,\mathbf{r})$ is defined by the equation \cite{damping}
\begin{equation}
\partial_t^2\alpha-c^2_s\nabla_{\mathbf{r}}^2\alpha+\frac{\nabla_{\mathbf{r}}^4\alpha}{4m^2}+\xi\partial_t\alpha=4\pi G\rho(\mathbf{r}-\mathbf{r}_\text{CM}(t)).
\label{perturbation-sphere}
\end{equation}
Here $c^2_s=\partial P/\partial n_0$ is the sound velocity squared connected with the ULDM self-interaction and $\xi$ is a parameter related to the damping term in the generalized Gross-Pitaevskii equation, which ensures that the system relaxes to equilibrium \cite{chavanis2019predictive}. The value of $\xi$ was estimated in \cite{chavanis2019predictive} via a generalized Einstein relation which expresses the fluctuation-dissipation theorem
\begin{equation}
\label{xi}
\xi=\frac{2T}{\hbar},    
\end{equation}
where $T$ is the effective temperature of the isothermal halo (we set the Boltzmann constant to the unity). The role of the damping term, whose magnitude is defined by $\xi$ for the dynamical friction force experienced by moving globular clusters in the Fornax dwarf galaxy was recently studied in \cite{damping}. 

In momentum space, for the considered circular steady-state motion, we find
\begin{equation}
\left(-\omega^2+c^2_s\mathbf{k}^2+\frac{\mathbf{k}^4}{4m^2}-i\xi\omega\right)\alpha(\omega,\mathbf{k})=4\pi G\!\!\int\limits^{+\infty}_{-\infty}\!\! dt\,\rho(\mathbf{k},t)\,e^{i\omega t-i\mathbf{k}\mathbf{r}_\text{CM}(t)},
\label{perturbation-sphere-momentum-1}
\end{equation}
where $\rho(\mathbf{k},t)$ is the Fourier transform of the mass density profile of the deformed Plummer sphere given by
\begin{equation}
\rho(\mathbf{k},t)=\rho\left(k_d(\mathbf{k},t)\right) =Mk_d(\mathbf{k},t)l_P\, K_1\left(k_d(\mathbf{k},t)l_P\right).
\label{density-momentum-2}
\end{equation}
Here $K_1(x)$ is the modified Bessel function of the second kind,
\begin{equation}
k_d(\mathbf{k},t)=\sqrt{k^2_z+a^2(k_{||}\cos\theta+k_t\sin\theta)^2+b^2(-k_{||}\sin\theta+k_t\cos\theta)^2},
\label{kd}
\end{equation}
and $k_{||}$ and $k_t$ are projections of $\mathbf{k}$ onto $\mathbf{r}_\text{CM}(t)$ and the tangential direction $\dot{\mathbf{r}}_\text{CM}(t)/|\dot{\mathbf{r}}_\text{CM}(t)|$, respectively.

According to the Poisson equation, the density inhomogeneity $\alpha(t,\mathbf{r})$ and the deformed Plummer sphere density $\rho(t,\mathbf{r})$ source a perturbation $\phi$ of the gravitational potential
\begin{equation}   \nabla_\mathbf{r}^2\phi(t,\mathbf{r})=4\pi G n_0(r)\alpha(t,\mathbf{r})+4\pi G\rho(\mathbf{ r}-\mathbf{r}_\text{CM}(t)).
\label{gravitational-potential}
\end{equation}
Solving Eq.~(\ref{perturbation-sphere-momentum-1}) for $\alpha(\omega,\mathbf{k})$, Eq.~(\ref{gravitational-potential}) gives the following gravitational potential due to perturbed ULDM density by moving deformed Plummer sphere (the complete perturbed gravitational potential is obviously $\phi=\phi_{\alpha}+\phi_{dPl}$, where $\phi_{dPl}$ is the Newtonian potential of the deformed Plummer sphere):
\begin{equation}
\phi_{\alpha}(t,\mathbf{r})=-4\pi G n_0 \int \frac{d\omega d\mathbf{k}}{(2\pi)^4}\frac{\alpha(\omega,\mathbf{k})}{k^2}\,e^{-i\omega t+i\mathbf{k}\mathbf{r}}.
\end{equation}
Then the local dynamical friction force density equals
\begin{multline}
\mathbf{f}_\text{fr}(t,\mathbf{r})=-\rho(\mathbf{r}-\mathbf{r}_\text{CM}(t))\nabla_{\mathbf{r}}\phi_{\alpha}(t,\mathbf{r})
=\\=4\pi G n_0 \rho(\mathbf{r}-\mathbf{r}_\text{CM}(t))\int \frac{d\omega d\mathbf{k}}{(2\pi)^4}\frac{i\mathbf{k}}{k^2}\alpha(\omega,\mathbf{k})e^{-i\omega t+i\mathbf{k}\mathbf{r}}=\\
=\frac{G^2n_0}{\pi^2} \rho(\mathbf{r}-\mathbf{r}_\text{CM}(t))\!\!\int\limits^{+\infty}_{-\infty}\!\!d t'\!\!\int\!\! d\omega d\mathbf{k}\frac{i\mathbf{k}}{k^2}\frac{\rho(\mathbf{k},t') e^{-i\omega (t-t')+i\mathbf{k}(\mathbf{r}-\mathbf{r}_\text{CM}(t'))}}{-\omega^2-i\xi\omega+c^2_sk^2+\frac{k^4}{4m^2}}.
\label{dynamical-force-local}
\end{multline}
It is convenient to change the variable $\tau=t-t'$ and then take into account that the integral over $\omega$ vanishes for $\tau<0$. We obtain
\begin{equation}
\mathbf{f}_\text{fr}(t,\mathbf{r})=\frac{G^2 n_0}{\pi^2} \rho(\mathbf{r}-\mathbf{r}_\text{CM}(t))\!\!\int\limits^{+\infty}_{0}\!\!d\tau \!\!\int\!\! d\omega d\mathbf{k}\frac{i\mathbf{k}}{k^2}\frac{\rho(\mathbf{k},t-\tau) e^{-i\omega\tau+i\mathbf{k}(\mathbf{r}-\mathbf{r}_\text{CM}(t-\tau))}}{-\omega^2-i\xi\omega+c^2_sk^2+\frac{k^4}{4m^2}}.
\label{dynamical-force-local-1}
\end{equation}
Clearly, this local dynamical force is different at different positions $\mathbf{r}$ within the deformed Plummer sphere producing a torque.

\color{black}
In the next section, we calculate the torque induced by the dynamical friction force.
\vspace{5mm}

\section{Torque caused by dynamical friction force}
\label{sec:torque}
\vspace{5mm}

Using Eq.~(\ref{dynamical-force-local-1}), we find that the total torque due to the dynamical friction force acting on a moving dwarf galaxy equals
$$
\mbox{\boldmath $\tau$} = \int\!\! d\mathbf{r}\,\mathbf{r}\times\mathbf{f}_\text{fr}(t,\mathbf{r}) \equiv \int\!\! d\mathbf{r}\,(\mathbf{r}-\mathbf{r}_\text{CM}(t))\times\mathbf{f}_\text{fr}(t,\mathbf{r})+\mathbf{r}_\text{CM}(t)\times\mathbf{F}_\text{fr}(t),
$$
where the last term $\mathbf{r}_\text{CM}(t)\times\mathbf{F}_\text{fr}(t)$ is the torque due to the total dynamical friction force which describes the loss of the angular moment of the deformed Plummer sphere in its orbital motion. The other term in the above equation defines the inner torque $\mbox{\boldmath $\tau$}_\text{inner}$ which does not have an analog in the case of point probe. It may cause rotation of the deformed Plummer sphere as a whole and, thus, is important for the study of internal kinematics of the modelled dwarf galaxies. This torque equals
$$
\mbox{\boldmath $\tau$}_\text{inner}(t)
=\frac{G^2n_0}{\pi^2}\!\!\int\!\! d\mathbf{r}\,\rho(\mathbf{r}-\mathbf{r}_\text{CM}(t))\!\!\int\limits^{+\infty}_{0}\!\!d\tau
$$
\begin{equation}
\times \int\!\! d\omega \frac{d\mathbf{k}}{k^2}\, i(\mathbf{r}-\mathbf{r}_\text{CM}(t))\times\mathbf{k} \frac{\rho(\mathbf{k},t-\tau)e^{-i\omega\tau+i\mathbf{k}\mathbf{r}-i\mathbf{k}\mathbf{r}_\text{CM}(t-\tau)}}{-\omega^2-i\xi\omega+c^2_sk^2+\frac{k^4}{4m^2}}.
\label{torque-local}
\end{equation}
Making the change of variable $\mathbf{r}=\mathbf{R}+\mathbf{r}_\text{CM}(t)$, we obtain
$$
\mbox{\boldmath $\tau$}_\text{inner}(t)
=\frac{G^2n_0}{\pi^2} \!\!\int\limits^{+\infty}_{0}\!\!d\tau \!\!\int\!\! d\omega \frac{d\mathbf{k}}{k^2}\,\frac{\rho(\mathbf{k},t-\tau)e^{-i\omega\tau+i\mathbf{k}\mathbf{r}_\text{CM}-i\mathbf{k}\mathbf{r}_\text{CM}(t-\tau)}}{-\omega^2-i\xi\omega+c^2_sk^2+\frac{k^4}{4m^2}}
$$
\begin{equation}
\times \int\!\! d\mathbf{R}\,e^{i\mathbf{k}\mathbf{R}}\rho(\mathbf{R},t)\,i\mathbf{R}\times\mathbf{k}.
\label{torque-local-1}
\end{equation}

The last integral in (\ref{torque-local-1}) can be rewritten in more simple form using 
\begin{equation}
\int\!\! d\mathbf{R}\,e^{i\mathbf{k}\mathbf{R}}\rho(\mathbf{R},t)\,i\mathbf{R}=\nabla_{\mathbf{k}}\!\int\!\! d\mathbf{R}\,e^{i\mathbf{k}\mathbf{R}}\rho(\mathbf{R},t)=\nabla_{\mathbf{k}}\rho(-\mathbf{k},t).
\label{gradk}
\end{equation}

In view of Eqs.~(\ref{density-momentum-2}) and (\ref{kd}), $\rho(-\mathbf{k},t)=\rho(\mathbf{k},t)$.
Further, we find in the cylinder coordinates
$$
\nabla_{\mathbf{k}}\rho\times\mathbf{k}=- \mathbf{e}_z\partial_{\phi}\rho+\mathbf{\hat{k}}_p\frac{k_z}{k_p}\partial_{\phi}\rho+\boldsymbol{\hat{\phi}}\left(k_p\partial_{k_z}\rho-k_z\partial_{k_p}\rho\right).
$$
Since $\rho$ depends on $k^2_z$, the last two terms in the cross product above are odd functions of $k_z$. Therefore, their contribution vanishes when integrating over $k_z$ in the expression for the inner torque. Thus, we conclude that as expected the inner torque has only one nonzero component in the direction perpendicular to the plane of orbital motion
$$
\mbox{\boldmath $\tau$}_\text{inner}(t)
=-\mathbf{e}_z\frac{G^2n_0}{\pi^2} \!\!\int\limits^{+\infty}_{0}\!\!d\tau \!\!\int\!\! d\omega \frac{d\mathbf{k}}{k^2}\,e^{-i\omega\tau+i\mathbf{k}\mathbf{r}_\text{CM}(t)-i\mathbf{k}\mathbf{r}_\text{CM}(t-\tau)}
$$
\begin{equation}
\times\,\frac{\rho(\mathbf{k},t-\tau)\partial_{\phi}\rho(\mathbf{k},t) }{-\omega^2-i\xi\omega+c^2_sk^2+\frac{k^4}{4m^2}}.
\label{torque-local-2}
\end{equation}

The poles of the integrand in Eq.~(\ref{torque-local-2}) are situated in the lower half-plane of the complex plane
\begin{equation}
\omega_{1,2}=-\frac{i\xi}{2} \pm D(k), \quad D(k)=\sqrt{-\frac{\xi^2}{4}+c^2_sk^2+\frac{k^4}{4m^2}}.
\label{poles}
\end{equation}
For $\tau>0$, closing the contour over $\omega$ in the lower half-plane and integrating over $\omega$ gives
$$
 \tau_\text{inner}(t)
=-\frac{2G^2n_0}{\pi} \!\!\int\limits^{+\infty}_{0}\!\!d\tau \!\!\int\!\! \frac{d\mathbf{k}}{k^2}\,e^{i\mathbf{k}\mathbf{r}_\text{CM}(t)-i\mathbf{k}\mathbf{r}_\text{CM}(t-\tau)}
$$
\begin{equation}
\times \frac{e^{-\xi\tau/2}}{D(k)}\sin\left(\tau D(k) \right)\, \rho(\mathbf{k},t-\tau)\partial_{\phi}\rho(\mathbf{k},t).
\label{torque-local-3}
\end{equation}
Further, using
$$
\mathbf{r}_\text{CM}(t)=r_0(\cos(\Omega t),\sin(\Omega t),0),\quad \mathbf{k}=(k_p\cos\phi,k_p\sin\phi,k_z),
$$
we find
$$
\mathbf{k}\mathbf{r}_\text{CM}(t)-\mathbf{k}\mathbf{r}_\text{CM}(t-\tau)=k_pr_0\big(\cos(\Omega t-\phi)-\cos(\Omega(t-\tau)-\phi)\big),
$$
\begin{equation}
k_{||}(t)=\mathbf{k}\cdot\frac{\mathbf{r}_\text{CM}(t)}{|\mathbf{r}_\text{CM}(t)|}=k_p\cos(\Omega t-\phi),\quad k_t(t)=\mathbf{k}\cdot\frac{\dot{\mathbf{r}}_\text{CM}(t)}{|\dot{\mathbf{r}}_\text{CM}(t)|}=-k_p\sin(\Omega t-\phi).
\label{relations}
\end{equation}
Then Eq.~(\ref{torque-local-3}) takes the form
$$
\tau_\text{inner}(t)
=-\frac{2G^2n_0}{\pi}\!\!\int\limits^{+\infty}_0\!\!d\tau  \!\!\int\limits^{+\infty}_{-\infty}\!\!dk_z\!\!\int\limits^{\infty}_0\!\!dk_p \frac{k_p}{k^2} \frac{e^{-\xi\tau/2}}{D(k)}\sin\left(\tau D(k) \right)
$$
\begin{equation}
\times \!\!\int\limits^{2\pi}_0\!\!d\phi\,e^{ik_pr_0\big(\cos(\Omega t-\phi)-\cos(\Omega(t-\tau)-\phi)\big)} \,\rho(\mathbf{k},t-\tau)\partial_{\phi}\rho(\mathbf{k},t),
\label{torque-local-4}
\end{equation}
where $\rho(\mathbf{k},t)$ is defined in Eq.(\ref{density-momentum-2}) and
$$
k_d(\mathbf{k},t)=\sqrt{k^2_z+a^2k^2_p\cos^2(\phi-\Omega t-\theta)+b^2k^2_p\sin^2(\phi-\Omega t-\theta)}.
$$

Changing the variable $\phi \to \phi+\Omega t$ in the last integral, we obtain
\begin{equation}
\int\limits^{2\pi-\Omega t}_{-\Omega t}\!\!d\phi\,e^{ik_pr_0\big(\cos\phi-\cos(\phi+\Omega\tau)\big)}\rho(k_d(\mathbf{k},-\tau))\partial_{\phi}\rho(k_d(\mathbf{k},0)).
\label{torque-local-5}
\end{equation}

Since the integral with respect to $\phi$ in Eq.~(\ref{torque-local-5}) is over the full period $2\pi$, this integral and hence the inner torque do not depend on $t$ as expected for the case of steady-state motion under consideration. Therefore, we set $t=0$ and the inner torque equals
$$
\tau_\text{inner}
=-\frac{2G^2n_0}{\pi}\!\!\int\limits^{+\infty}_0\!\!d\tau \!\!\int\limits^{+\infty}_{-\infty}\!\!dk_z\!\!\int\limits^{\infty}_0\!\!dk_p\frac{k_p}{k^2} \frac{e^{-\xi\tau/2}}{D(k)}\sin\left(\tau D(k) \right)
$$
\begin{equation}
\times\!\!\int\limits^{2\pi}_0\!\!d\phi\,e^{ik_pr_0\big(\cos\phi-\cos(\phi+\Omega\tau)\big)}\, \rho(k_d(\mathbf{k},-\tau)) \partial_{\phi}\rho(k_d(\mathbf{k},0)).
\label{torque-local-6}
\end{equation}

Let us rewrite the inner torque (\ref{torque-local-6}) in an explicitly real form. For this, we split the integral over $\phi$ in two equal parts $I_{\phi}=I_{\phi}/2+I_{\phi}/2$ and make the change of variable $\phi \to \phi +\pi$ in the second term which gives $\cos\phi-\cos(\phi+\Omega\tau) \to -(\cos\phi-\cos(\phi+\Omega\tau))$. Taking into account that we integrate over the full period, the limit of integration in the second term can be kept unchanged. Then Eq.~(\ref{torque-local-6}) takes the explicitly real form
$$
\tau_\text{inner}
=-\frac{2G^2n_0}{\pi}\!\!\int\limits^{+\infty}_0\!\!d\tau \!\!\int\limits^{+\infty}_{-\infty}\!\!dk_z\!\!\int\limits^{\infty}_0\!\!dk_p\frac{k_p}{k^2} \frac{e^{-\xi\tau/2}}{D(k)}\sin\left(\tau D(k) \right) 
$$
\begin{equation}
\times \!\!\int\limits^{2\pi}_0\!\!d\phi\cos\left[k_pr_0\big(\cos\phi-\cos(\phi+\Omega\tau)\big)\right]\,\rho(k_d(\mathbf{k},-\tau))\partial_{\phi}\rho(k_d(\mathbf{k},0)).
\label{torque-local-7}
\end{equation}
Using
\begin{equation}
\partial_{\phi}\rho(k_d)=\partial_{\phi}\left[ M k_d l_P\, K_1(k_dl_P)\right]=-Mk_dl_p^2 K_0(k_d l_p) \partial_{\phi} k_d,  
\end{equation}
where
$$
\partial_{\phi} k_d(\mathbf{k},0)=\frac{k_p^2}{2k_d(\mathbf{k},0)}(b^2-a^2)\sin(2\phi-2\theta),
$$
the inner torque takes the following final form:
\begin{multline}
\tau_\text{inner}
=(b^2-a^2)\frac{G^2M^2l_p^3n_0}{\pi}\!\!\int\limits^{+\infty}_0\!\!d\tau \!\!\int\limits^{+\infty}_{-\infty}\!\!dk_z\!\!\int\limits^{\infty}_0\!\!dk_p\frac{k_p^3}{k^2}\frac{e^{-\xi\tau/2}}{D(k)}  \sin\left(\tau D(k) \right)\\
\times\!\!\int\limits^{2\pi}_0\!\! d\phi\cos\left[k_pr_0\big(\cos\phi-\cos(\phi+\Omega\tau)\big)\right]\,k_d(\mathbf{k},-\tau)\\
\times K_1(l_pk_d(\mathbf{k},-\tau)) K_0(l_pk_d(\mathbf{k},0)) \sin(2\phi-2\theta).
\label{torque-local-10}
\end{multline}
Obviously, this torque vanishes for $a=b$ when the dwarf galaxy is an ellipsoid of revolution with respect to the axis $z$.

If this inner torque is not compensated, it will induce rotation of the dwarf galaxy under consideration. Further, for dwarf galaxies orbiting the Milky Way, there exists a torque due to tidal bulges displaced by angle $\theta$ from the axis defined by $\mathbf{r}_\text{CM}(t)$ as we assumed in the setup of our model. This means that we should calculate and take into account the tidal torque.

\section{Torque due to tidal forces}
\label{sec:torque-tidal}
\vspace{5mm}

The tidal torque is defined by the equation
\begin{equation}
\mbox{\boldmath $\tau$}_g = \int\!\! d\mathbf{r}\,(\mathbf{r}-\mathbf{r}_\text{CM}(t))\times\mathbf{f}_g(t,\mathbf{r}),
\label{torque-tidal}
\end{equation}
where $\mathbf{f}_g(t,\mathbf{r})$ is the local gravitational force density acting on dwarf galaxy orbiting the Milky Way
$$
\mathbf{f}_g(t,\mathbf{r})=-\rho(\mathbf{r}-\mathbf{r}_\text{CM}(t))\nabla_{\mathbf{r}}\phi_\text{MW}(t,\mathbf{r}),\qquad \phi_\text{MW}(t,\mathbf{r})=-\frac{GM_\text{MW}}{r}.
$$
Equation (\ref{torque-tidal}) gives
$$
\mbox{\boldmath $\tau$}_g = -GM_\text{MW}\!
\!\int\!\! \frac{d\mathbf{r}}{r^3}\,\rho(\mathbf{r}-\mathbf{r}_\text{CM}(t))\,\,(\mathbf{r}-\mathbf{r}_\text{CM}(t))\times\mathbf{r}
$$
$$
=-GM_\text{MW}\!\!\int\!\! \frac{d\mathbf{r}}{r^3}\,\rho(\mathbf{r}-\mathbf{r}_\text{CM}(t))\,\,(\mathbf{r}-\mathbf{r}_\text{CM}(t))\times\mathbf{r}_\text{CM}(t)
$$
\begin{equation}
=-GM_\text{MW}\!\int\! \frac{d\mathbf{R}\,\,\rho(\mathbf{R})}{|\mathbf{R}+\mathbf{r}_\text{CM}(t)|^3}\,\,\,\mathbf{R}\times\mathbf{r}_\text{CM}(t).
\label{torque-tidal-1}
\end{equation}
Since $\mathbf{r}_\text{CM}(t)$ lies in the equatorial plane and $\rho(\mathbf{R})$ is even in $R_z$ (see, Eq.~(\ref{Plummer-sphere-deformed})), torque (\ref{torque-tidal-1}) reduces to
\begin{multline}
\mbox{\boldmath $\tau$}_g=-GM_\text{MW}\!\int\! \frac{dR_zdR_{||}dR_t\,\,\rho(\mathbf{R})}{(R^2_z+R^2_t+(R_{||}+r_\text{CM}(t))^2)^{3/2}}\,\,\,\mathbf{R}_t\times\mathbf{r}_\text{CM}(t)\\
=\mathbf{e}_zGM_\text{MW} r_0 \!\int\! \frac{dR_zdR_{||}dR_t\,\,\rho(\mathbf{R}) R_t }{(R^2_z+R^2_t+(R_{||}+r_\text{CM}(t))^2)^{3/2}} ,
\label{torque-tidal-2}
\end{multline}
which is like the inner torque $\mbox{\boldmath $\tau$}_\text{inner}$ given by Eq.(\ref{torque-local-2}) is directed perpendicular to the plane of orbital motion. Finally, note that if angle $\theta$ in Eq.~(\ref{Plummer-sphere-deformed}) is zero, then the deformed Plummer sphere density profile $\rho(\mathbf{R})$ is even in $R_t$ and, therefore, torque $\mbox{\boldmath $\tau$}_g$ vanishes when integrated over $R_t$.

In dimensionless variables $X=R_{||}/l_P$, $Y=R_t/l_P$, and $Z=R_z/l_P$, which are useful for numerical studies, the density profile (\ref{Plummer-sphere-deformed}) takes the form
\begin{equation}
\rho(\mathbf{R})=\frac{3M }{4\pi l_P^3 ab}\frac{1}{\left(Z^2+A^2\right)^{5/2}},
\label{Plummer-sphere-deformed2}
\end{equation}
where
\begin{equation*}
A^2=1+\frac{(X\cos\theta+Y\sin\theta)^2}{a^2}+\frac{(-X\sin\theta+Y\cos\theta)^2}{b^2},   
\end{equation*}
and the tidal torque equals
\begin{equation}
\mbox{\boldmath $\tau$}_g=
 \mathbf{e}_z  r_0  \frac{3GM M_\text{MW}}{4\pi l_P^2 ab} \!\!\int\!\! dXdY Y\!\!\int\limits_{-\infty}^{\infty}\!\! \frac{d Z }{(Z^2+B^2)^{3/2} \left(Z^2+A^2\right)^{5/2}}
\label{torque-tidal-3}
\end{equation}
with $B^2=Y^2+(X+r_0/l_p)^2$.

The integral over $Z$ in Eq.(\ref{torque-tidal-3}) can be calculated analytically. Changing the variable $Z=A\tan\beta$, we find for the integral over $Z$
\begin{equation}
    I_z=\frac{2}{A^4B^3}\!\int\limits_0^{\pi/2}\!\frac{\cos^6\beta d\beta}{\left(1-\frac{B^2-A^2}{B^2}\sin^2\beta \right)^{3/2}}.
\end{equation}
Denoting $(B^2-A^2)/B^2\equiv k^2$ and once again changing the integration variable $\sin\beta=x$, we obtain
\begin{equation}
    I_z=\frac{2}{A^4B^3}\int\limits_0^1\frac{(1-x^2)^{5/2} dx}{\left(1-k^2x^2\right)^{3/2}}=\frac{2\left((9k^4-17k^2+8)K(k)-(3k^4-13k^2+8)E(k) \right)}{3A^4B^3k^6},
\end{equation}
where $K(k)$  and $E(k)$ are the complete elliptic integrals of the first and second kind \cite{Gradstein}.

Finally, we find the following tidal torque:
\begin{equation}
\label{tau_g}
\tau_g =\frac{3GM M_\text{MW}}{4\pi l_P   } T_g,   
\end{equation}
where $T_g$ is dimensionless tidal torque
\begin{equation}
 T_g =\frac{1}{abL_p} \int\!\! \int dX dY\, Y I_z(X,Y) \label{T}  
\end{equation}
with $L_p=l_p/r_0$.

\section{Numerical results for torques of dwarf galaxies orbiting the Milky Way}
\label{sec:numerical}

To proceed with numerical analysis, it is convenient to rewrite the inner torque (\ref{torque-local-10}) also in dimensionless variables $K_z=k_zr_0$, $K_p=k_p r_0$, and $\mathcal{T}=\tau c_s/r_0$. Other dimensionless parameters are the Mach number $\mathcal{M}=r_0\Omega/c_s$, $A=mr_0c_s$, and $\Xi=\xi r_0/c_s$. Then
$$
  D(k)=\frac{\tilde{D}(K)}{2mr_0^2}, \quad \tilde{D}(K)=\sqrt{K^4+4A^2K^2-A^2\Xi^2}
$$
and the inner torque (\ref{torque-local-10}) takes the following form in dimensionless variables:
\begin{equation}
\mbox{\boldmath $\tau$}_\text{inner}
=\mathbf{e}_z\frac{4\pi G^2M^2n_0 r_0}{c_s^2} T_\text{inner},
\label{inner-torque}
\end{equation}
$$
T_\text{inner}=(b^2-a^2)\frac{2A L_p^3}{\pi^2}\!\!\int\limits^{+\infty}_0\!\!d\mathcal{T} \!\!\int\limits^{+\infty}_{0}\!\!dK_z\!\!\int\limits^{\infty}_0\!\!dK_p\frac{K_p^3}{K^2}\frac{e^{-\Xi \mathcal{T}/2}}{\tilde{D}(K)} \sin\bigg(\frac{\mathcal{T}\tilde{D}(K)}{2A} \bigg) 
$$
$$
\times\!\!\int\limits^{\pi}_0\!\! d\phi\cos\left[K_p\big(\cos\phi-\cos(\phi+\mathcal{M}\mathcal{T})\big)\right]\,K_d(\mathbf{K},-\mathcal{T}) 
$$
\begin{equation}
\times K_1(L_pK_d(\mathbf{K},-\mathcal{T})) K_0(L_pK_d(\mathbf{K},0)) \sin(2\phi-2\theta).
\label{torque-local-11}
\end{equation}

Since the inner torque (\ref{torque-local-11}) is proportional to $a^2-b^2$, its value is largest for dwarf galaxies with considerable ellipticity $\varepsilon=1-b/a$. Further, our model is developed for dwarf galaxies moving around the Milky Way on circular orbits. Hence, we choose for our analysis dwarf galaxies listed in Table \ref{tab:1} whose orbits have low eccentricity $e$. The last three columns of this table define the values of the Mach number $\mathcal{M}=v_\text{tan}/c_s$ and parameters $A=mr_0c_s/\hbar$ and $\Xi=\xi r_0/c_s$.

\begin{table}[htb]
 \caption{Galactocentric distance $r_\text{GC}$, galactocentric tangential velocity $v_\text{tan}$, and eccentricity $e$ of dwarf galaxies orbiting the Milky Way in the PNFW model \cite{Gaia} with ellipticity $\varepsilon=1-b/a$ in the Plummer model \cite{Munoz} with radius $l_p$.}
 \vspace{4mm}
    \centering
    \begin{tabular}{c|c|c|c|c|c|c|c|c|c}
     dwarf &$r_\text{GC}$ & $v_\text{tan}$ & $e$ & $\varepsilon$ & $l_p$ & $r_\text{GC}/l_p$ &$\mathcal{M}$&$A$ & $\Xi$ \\ galaxy
    & (kpc)  & (km s$^{-1}$) & & & (pc) &  & & \\ \hline
    CarI & 107.6  &192.2 & 0.3  &  0.36 & 308 & 349 &1.69& 1912& 3823 \\ 
    FnxI & 141.0  &126.9 & 0.26 &  0.29 & 838 & 168 & 1.12 & 2505& 5010 \\ 
    SclI & 86.0  & 162.7 & 0.36 &  0.33 &  280 & 307 & 1.43 & 1528& 3056 \\ 
    UMaI & 102.0 &133.3 & 0.25 & 0.59 & 234 & 436 & 1.17 & 1812& 3624 \\
    UMiI & 77.8 &147.6  & 0.37 & 0.55 & 407 & 191 & 1.30 & 1382& 2764 \\
    WilI & 49.3  &159.2 & 0.25 & 0.47 & 27.7 & 1780  & 1.40 &876& 1752 \\
    \end{tabular}
    \label{tab:1}
\end{table}

The ULDM pressure $P = K\rho^2 + \rho k_B T / m$ in the isothermal halo \cite{chavanis2019predictive} is dominated by the thermal term $P = \rho k_B T / m$ with the sound velocity given by $c_s^2 = \partial P / \partial \rho = k_B T / m$. For the ULDM effective temperature in the Milky Way halo $T = 4.3 \times 10^{-29}\,K$ and $m = 3 \times 10^{-22}\, \text{eV}$ \cite{Korshynska}, we obtain $c_s \approx 114$ km~s$^{-1}$. Note that the Mach number for the dwarf galaxies under consideration is $\mathcal{M} > 1$, i.e., the orbital motion of the dwarf galaxies proceeds in the supersonic regime. For the damping term, we have
\begin{equation}
  \xi=\frac{2T}{\hbar}=1.3\times10^{-13} \text{ s}^{-1}.
\end{equation}

\begin{figure}[htb]
    \centering    \includegraphics[width=0.6\linewidth]{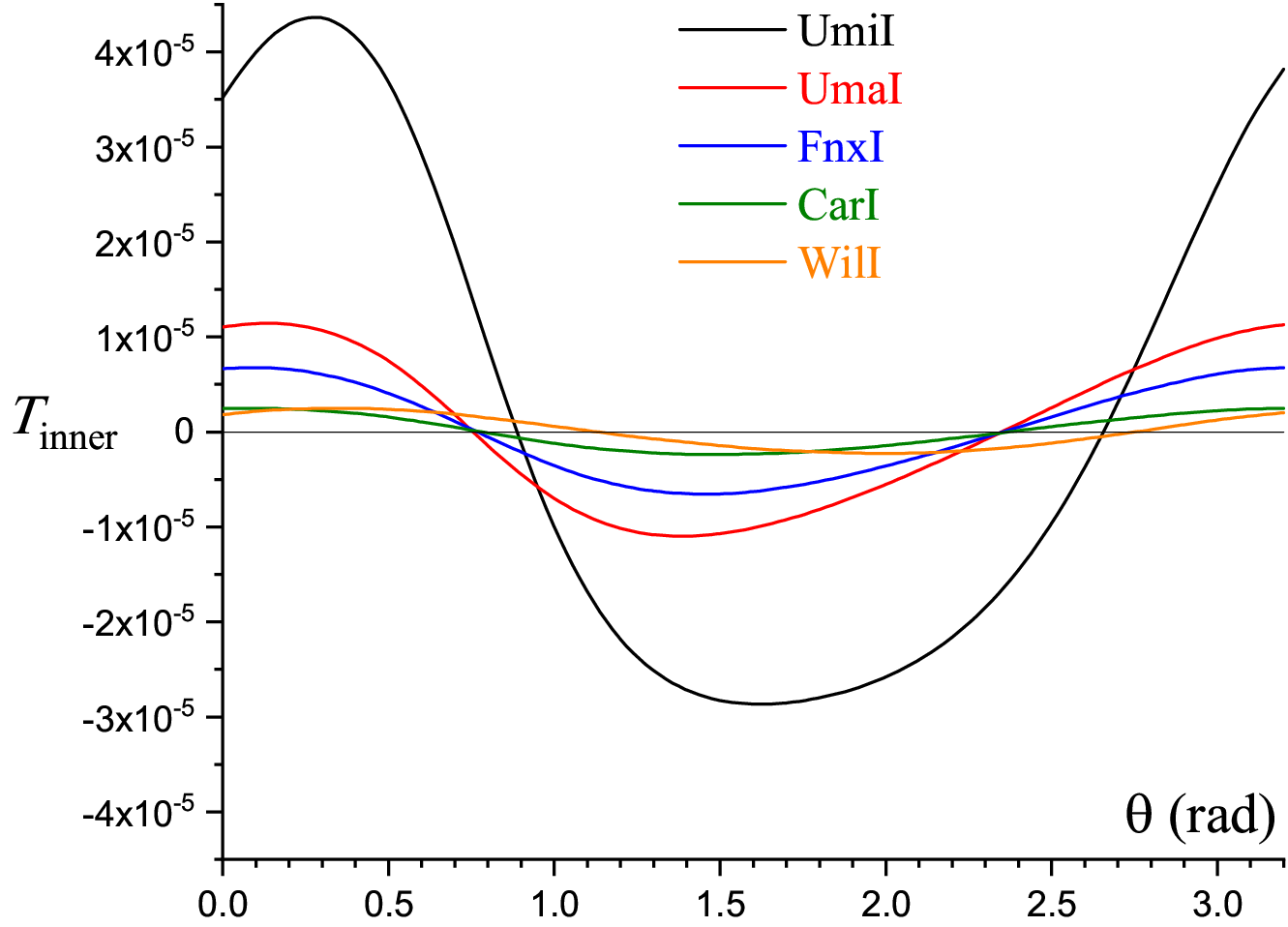}
    \caption{Dimensionless inner torque $T_\text{inner}$ as a function of angle $\theta$ for dwarf galaxies UmiI, UmaI, FnxI, CarI, and WilI.}
    \label{fig_Tinner_theta2}
\end{figure}

We plot in Figure~\ref{fig_Tinner_theta2} the dimensionless inner torque (\ref{torque-local-11}) as a function of angle $\theta$. One can see that this dependence is non-monotonous and changes sign. In addition, the torque vanishes for different values of $\theta$ for different dwarf galaxies.

\begin{figure}[htb]
    \centering
\includegraphics[width=0.5\linewidth]{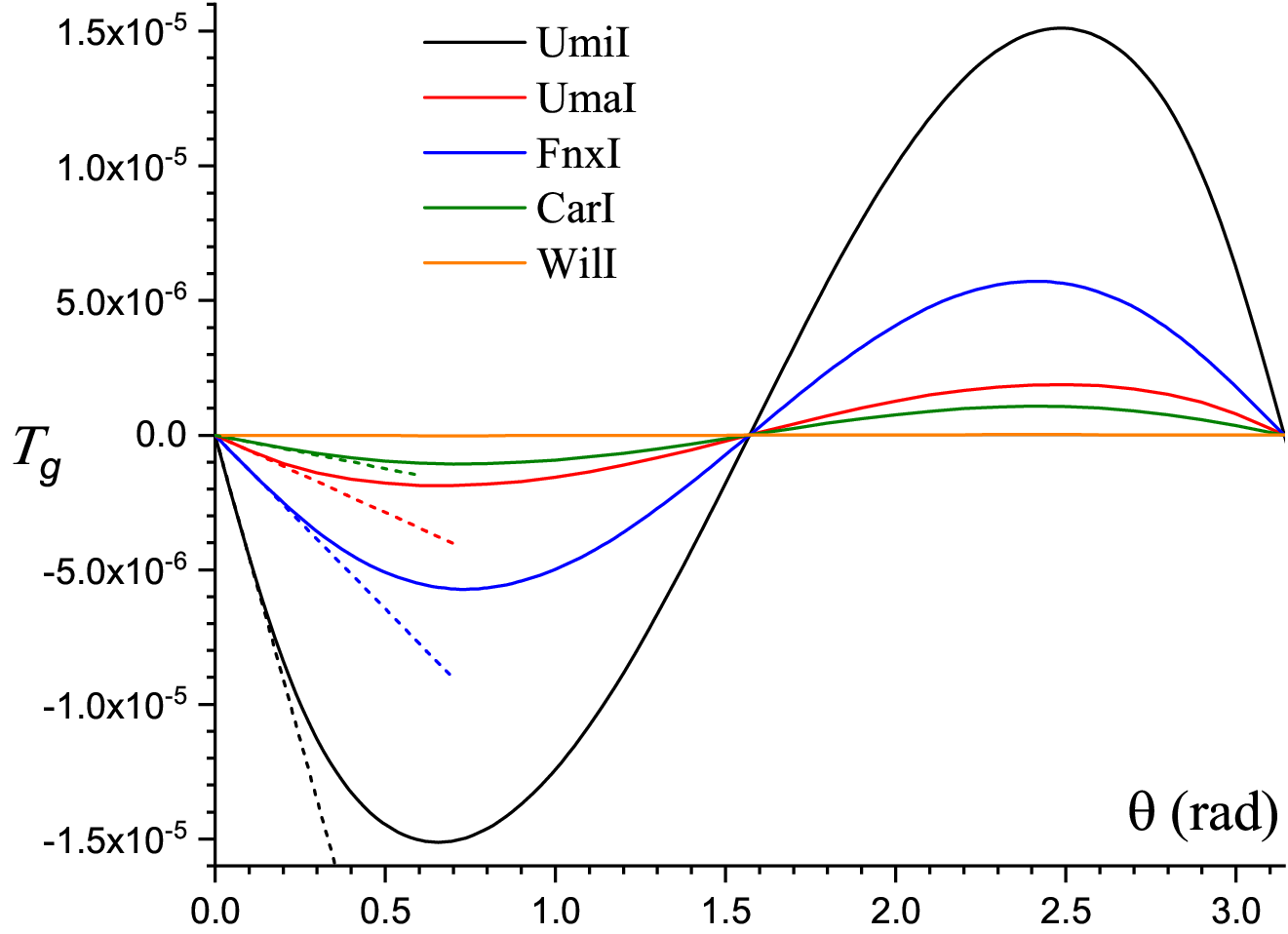}
\includegraphics[width=0.46\linewidth]{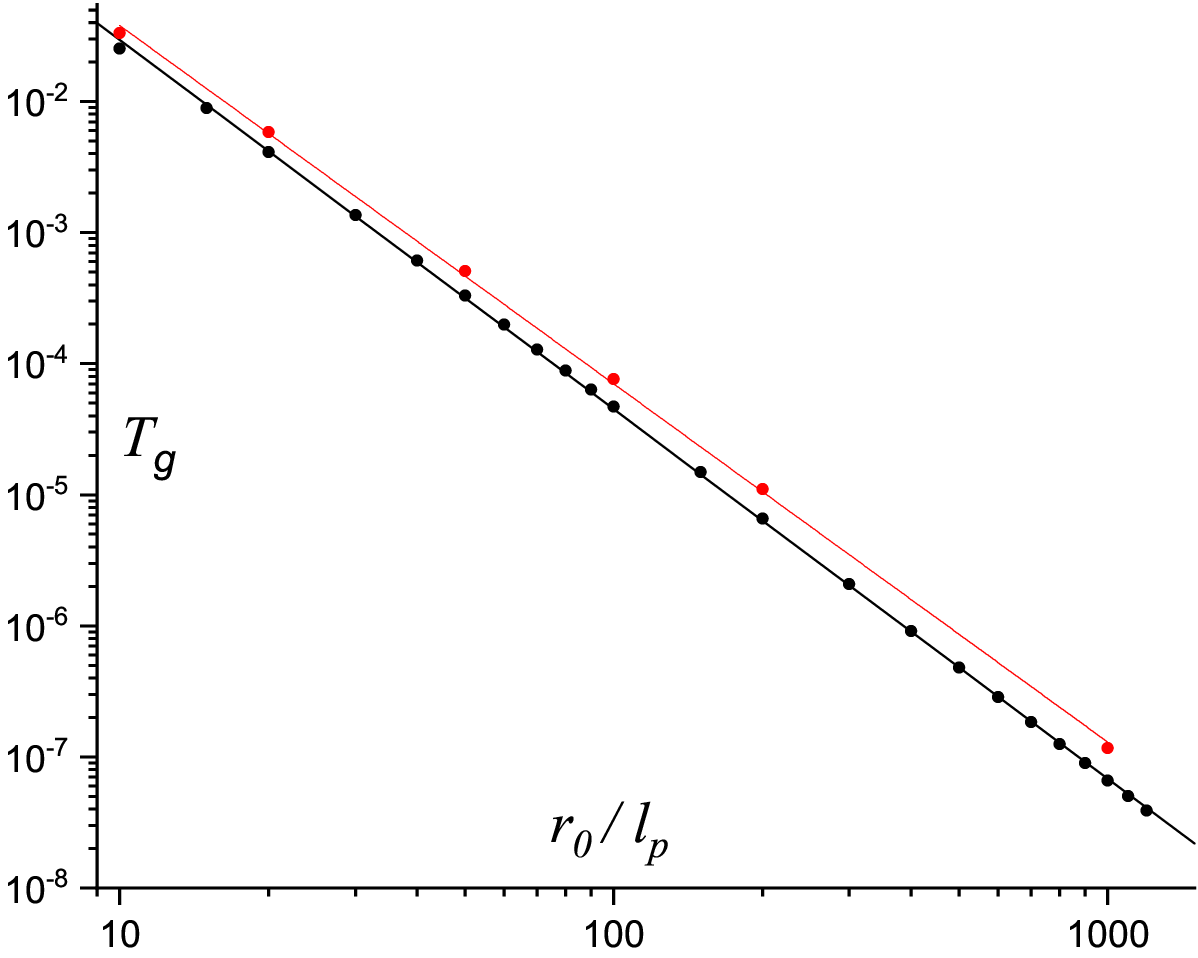}
    \caption{Left panel: Dimensionless tidal torque $T_g$ as a function of $\theta$ for dwarf galaxies UmiI, UmaI, FnxI, CarI, and WilI. Dashed lines correspond to the linear approximation $T_g=-\alpha\,\theta$ with values of $\alpha$ listed in Table~\ref{tab:2}. Right panel: Dimensionless tidal torque $T$ as a function of $r_0/l_p$ for $a=2$, $b=0.8$, $\theta=0.2$ rad (black solid line) and $\theta=0.5$ rad (red dashed line) with points defined numerically using Eq.~(\ref{T}) and the straight lines are linear approximations.}
    \label{fig_T_theta}
\end{figure}

Calculating the dimensionless tidal torque $T_g$ given by Eq.(\ref{T}) numerically, we depict it as a function of $\theta$ in the left panel of Fig.~\ref{fig_T_theta}. As to the dependence of $T_g$ on $r_0/l_p$ in a log-log plot is well approximated by a linear function and is shown in the right panel of Fig.~\ref{fig_T_theta}, i.e., $T_g \approx 10^{1.289}(r_0/l_p)^{-2.818}$. One can see from the left panel in  Fig.~\ref{fig_T_theta} that the tidal torque is negative for $\theta\in  (0,\pi/2)$, i.e., it is directed opposite to the $z$-axis and tends to decrease the angle $\theta$. $T_g$ reaches its maximum absolute value for $\theta\approx36\degree$ and vanishes at $\theta=\pi/2$. However, the latter position is unstable because a small change $\delta\theta$ leads to the appearance of a torque which further increases $\delta\theta$. For $\theta\in  (\pi/2,\pi)$, the tidal torque is positive and tends to increase $\theta$ up to $\pi$.


Having determined dimensionless torques, we compare the inner torque (\ref{inner-torque}) and the tidal torque (\ref{tau_g}) by calculating their ratio
\begin{equation}
    \frac{\tau_\text{inner}}{\tau_g}=k \frac{T_\text{inner}}{T_g}, \quad k=\frac{(4\pi)^2 G M n_0(r_0) r_0 l_P}{3 M_\text{MW} c_s^2}.
\end{equation}
Using the Navarro–Frenk–White profile for the ULDM density in the vicinity of the Milky Way
\begin{equation}
    n_0(r)=\frac{n_s}{(r/r_s)(1+r/r_s)^2},
\end{equation}
with $r_s=16$ kpc, $n_s=8.4868\times10^6$ $M_\odot/\text{kpc}^3$, and $M_\text{MW}=8\times10^{11} M_\odot$ \cite{Bovy},
we easily find
\begin{equation}
k=\frac{2.9728\times10^{-12} M[M_\odot] l_P[\text{kpc}]}{(1 + r_0[\text{kpc}]/16)^2}.  
\end{equation}
Hence, coefficient $k$ lies in the range $1.28\times10^{-14} M $ for UmaI to $4.94\times10^{-14}M$ for WilI. Taking the estimates of total (baryonic plus dark matter) mass of dwarf galaxies $M\sim10^8 M_\odot$, we conclude that the ratio of the inner torque to the tidal torque is negligible. However, this ratio can be substantial for objects of similar mass rotating around their common center of mass or in the process of merger. Then the inner torque should be taken into account in the analysis.


\section{Moment of inertia and period of oscillations}
\label{sec:period}

The form of the tidal torque in the left panel of Fig.\ref{fig_T_theta} suggests the possibility of oscillations of dwarf galaxies around their equilibrium position. To describe these oscillations, we need to determine the moment of inertia of dwarf galaxies. To proceed and obtain an estimate of the moment of inertia, we assume that oscillations or full scale rotation do not affect the matter distribution profile. Of course, to check this assumption a coupled system of equations which describe dwarf galaxy matter and perturbed Milky Way halo should be analyzed and solved. We leave this problem for future studies. Then taking this assumption into account, the relevant for us $z$-component is given by the standard expression
\begin{equation}
I_z=\iiint \rho(\mathbf{r})(x^2+y^2)d\mathbf{r},
\end{equation}
where
$$
\rho(\mathbf{r})=\frac{3M}{4\pi l^3_Pab}\frac{1}{\left(1+\frac{z^2}{l^2_p}+\frac{x^2}{a^2l^2_P}+\frac{y^2}{b^2l^2_P}\right)^{5/2}}.
$$
Integrating over $x$ and $y$ gives a logarithmically divergent integral
\begin{equation}
I_z=M l_P^2\frac{a^2+b^2}{2}\!\!\int\limits_{-\infty}^{\infty}\!\! \frac{dz}{\sqrt{1+z^2}}.
\end{equation}
This divergence is regulated by a certain physical scale related to the size of dwarf galaxies. Since the logarithmic dependence is quite mild, the exact value of this physical scale is not important and could approximate the moment of inertia as $I_z=CMl_P^2$, where $C \sim O(1)$. Since the tidal torque near the equilibrium position $\theta=0$ is given by $T_g=-\alpha\theta$, harmonic
oscillations for small deviations $\theta$ are realized with frequency
$$
\omega=\sqrt{\frac{\tau_g/\theta}{I_z}}=\sqrt{\frac{3G M_\text{MW}\,\alpha}{4\pi C l_P^3 }}
$$
\begin{equation}
=\sqrt{\frac{3 \times4.30091\times10^{-6}[\text{ kpc (km/s)}^2 / M_{\odot}] M_\text{MW}[M_\odot]\,\alpha}{4\pi C l_P[\text{kpc}] (l_p[\text{km}])^2 }}.
\label{frequency}
\end{equation}

For the dwarf galaxy UmiI with $l_p=407$ pc and $C=5$, we find $\omega\approx10.5$ Gyr$^{-1}$ and the period of oscillations $2\pi/\omega\approx 0.60$ Gyr.
For other dwarf galaxies, the frequency and period of their oscillations are given in the 3rd and 4th columns of Table~\ref{tab:2}.

The characteristic velocity associated with oscillations of dwarf galaxies can be estimated as $\omega l_p$. For the dwarf galaxy CarI, the obtained value $\omega l_p = 1.15$ km/s is consistent with the change in the mean radial velocity of 2–3 km/s reported in \cite{Fabrizio}. Still a higher tangential velocity of $|V_T| = 9.6 \pm 4.5$ km/s was reported in \cite{Watkins}. For the dwarf galaxy UMiI, the tangential velocity is the lowest among all considered galaxies, as the last column in Table~\ref{tab:2} implies.

\begin{table}[htb]
 \caption{The values of parameter $\alpha$ in the linear approximation for the tidal torque $T_g=-\alpha\,\theta$, the frequency of oscillations $\omega$, the period of oscillations $T$, and the characteristic velocity $\omega l_p$.}
    \centering
    \begin{tabular}{c|c|c|c|c}
     dwarf&$\alpha\cdot10^5$ &$\omega$ & $T$ & $\omega l_p$\\
    galaxy& & (Gyr$^{-1}$) & (Gyr)& (km/s)\\ \hline
    CarI & 0.248 & 3.73 & 1.69 & 1.15\\ 
    FnxI & 1.29 & 1.89 & 3.32  & 1.59\\ 
    SclI & 0.296 & 4.71 &  1.33 & 1.32 \\ 
    UMaI&  0.575& 8.56 & 0.734 & 2.01\\ 
    UMiI& 4.53 & 10.5  & 0.600 & 4.28\\ 
    WilI& $4.59\cdot10^{-3}$ & 18.8 & 0.335 & 0.52\\ 
    \end{tabular}
    \label{tab:2}
\end{table}

\section{Discussion and conclusions}
\label{sec:Conclusion}

Modeling dwarf galaxies as deformed Plummer spheres, the dynamical friction force and the corresponding induced torque for dwarf galaxies orbiting the Milky Way are determined in ultralight dark matter models. In addition, the torque due to the tidal force of the Milky Way is calculated. Numerically solving the obtained analytic formulae, both torques are found for CarI, FrxI, UMaI, UMiI, and WilI dwarf galaxies. One of our main findings is that for all considered dwarf galaxies the torque produced by the dynamical friction force is suppressed by three or four orders of magnitude compared to the tidal torque. Thus, to study the internal kinematics of dwarf galaxies orbiting much more massive galaxies, it suffices to retain only the tidal torque in the corresponding analysis. However, the inner torque can be substantial for objects of similar mass rotating around their common center of mass or in the process of merger. Then this torque should be included in the analysis.

As to astrophysical observations of internal kinematics of dwarf galaxies, they show that many of the Milky Way’s classical dwarf spheroidals are dispersion dominated, i.e, $(|V_{\text{rot}}|/\sigma_{v} \leq 1)$ with the Carina dwarf spheroidal being the exception \cite{Watkins}. According to the study of dwarf satellites of MV/M31-like hosts from the IllustrisTNG50 simulation \cite{Aparicio}, internal rotation tends to be stronger in less evolved dwarfs, or those that are less tidally stripped. As satellites pass close to the Milky Way center (pericenters), tidal forces tend to reduce their internal ordered rotation, converting it into random motions in agreement with simulations \cite{Lokas}. Gaia (especially its EDR3) has enabled for the first time detecting proper motions of stars within dwarf galaxies, which allows measurement of sky‑plane rotation (not just line‑of‑sight velocity component) \cite{Watkins}.

According to our study, the tidal torque is characterized by a non-monotonous dependence on angle $\theta$ which defines misalignment of the largest axis of a dwarf galaxy with respect to the radius-vector of its center-of-mass position. For small $\theta$, it could be approximated by a linear dependence. For misaligned dwarf galaxies, this implies the possibility of oscillations around their equilibrium position. We found that the period of oscillations lies in a large interval from 0.335 Gyr for WilI to 3.32 Gyr for FnxI. As to typical velocities connected with dwarf galaxies oscillations, they range from 0.52 km/s for WilI to 4.28 km/s for UMiI.

Obviously, our assumption of a solid body rotation may be unrealistic for dwarf galaxies. Therefore, its validity should be checked in future studies of the internal kinematics of dwarf galaxies orbiting the Milky Way. In addition, it would be definitely interesting to extend our study of internal kinematics, restricted here to the analysis of small oscillations of dwarf galaxies to the case of full-scalele rotation which is astrophysically observed in dwarf spiral and irregular galaxies \cite{Flores,Bosma,Oh,Read,Dehghani,Camalich}.

\vspace{5mm}

\centerline{\bf Acknowledgements}
\vspace{5mm}

The authors are grateful to B.I. Hnatyk, Y. Revaz, and A.I. Yakimenko for fruitful and stimulating discussions and acknowledge support from the SNSF through the Swiss-Ukrainian Joint research project "Cosmic waltz of baryonic and ultralight dark matter: interaction and dynamical interplay" (grant No. IZURZ2{\_}224972).

\end{document}